\documentclass[english]{revtex4-2}
\usepackage[T1]{fontenc}
\usepackage[latin9]{inputenc}
\usepackage{babel}
\usepackage{amstext}
\usepackage{graphicx}
\usepackage[unicode=true,pdfusetitle,
 bookmarks=true,bookmarksnumbered=false,bookmarksopen=false,
 breaklinks=false,pdfborder={0 0 1},backref=false,colorlinks=false]
 {hyperref}

\makeatletter

\providecommand{\tabularnewline}{\\}

\makeatother

\begin{document}
\title{Ising model on a \emph{restricted} scale-free network}
\author{R. A. Dumer }
\email{rafaeldumer@fisica.ufmt.br}

\affiliation{Instituto de Física - Universidade Federal de Mato Grosso, 78060-900,
Cuiabá, Mato Grosso, Brazil.}
\author{M. Godoy}
\email{mgodoy@fisica.ufmt.br}

\affiliation{Instituto de Física - Universidade Federal de Mato Grosso, 78060-900,
Cuiabá, Mato Grosso, Brazil.}
\begin{abstract}
The Ising model on a \emph{restricted} scale-free network (SFN) has
been studied employing Monte Carlo simulations. This network is described
by a power-law degree distribution in the form $P(k)\sim k^{-\alpha}$,
and is called restricted, because independently of the network size,
we always have fixed the maximum $k_{m}$ and a minimum $k_{0}$ degree
on distribution, being that for it, we only limit the minimum network
size of the system. We calculated the thermodynamic quantities of
the system, such as, the magnetization per spin $\textrm{m}_{\textrm{L}}$,
the magnetic susceptibility $\textrm{\ensuremath{\chi}}_{\textrm{L}}$,
and the reduced fourth-order Binder cumulant $\textrm{U}_{\textrm{L}}$,
as a function of temperature $T$ for several values of lattice size
$N$ and exponent $1\le\alpha\le5$. For the values of $\alpha$,
we have obtained the finite critical points due to we also have finite
second and fourth moments in the degree distribution, and the phase
diagram was constructed for the equilibrium states of the model in
the plane $T$ versus $k_{0}$, $k_{m}$, and $\alpha$, showing a
transition between the ferromagnetic $F$ to paramagnetic $P$ phases.
Using the finite-size scaling (FSS) theory, we also have obtained
the critical exponents for the system, and a mean-field critical behavior
is observed.
\end{abstract}
\maketitle

\section{Introduction}

Hyperlinks pointing from one web page to another (World Wide Web),
computers physically linked (internet), actors that have acted in
a movie together, scientists that have an article together, and proteins
that bind together experimentally, are some of the cases that, when
analyzed in terms of nodes and edges of a network, they are part of
a broad group of real systems in which the degree distribution has
a power-law tail \citep{1,2}. This degree distribution has the form
$P(k)\sim k^{-\alpha}$, representing the probability of a site in
the network to have a degree $k$, i.e., $k$ edges linked to it,
with exponent $\alpha$. Barabási-Albert \citep{3} proposed a growing
process of network creation, in which the degree distribution resultant
also has the form of power-law. That growing process is based only
on two generic mechanisms: (i) networks expand continuously by the
addition of new vertices, and (ii) new vertices attach preferentially
to sites that are already well connected. With these mechanisms, most
connected sites are most likely to receive new connections, and a
``rich-get-richer'', and self-organization phenomenon, as in real
networks, is observed.

Due to the applicability of the networks with a power-law degree distribution,
also called scale-free networks, it was been implemented in many physical
problems \citep{4,5,6,7,8,9,10,11}. Highlighting the critical phenomena
arising from the Ising model, from an analytical way, Dorogovtsev,
Goltsev, and Mendes showed that its critical behavior is very dependent
on the distribution of connections \citep{12,13}. From this dependence,
when $\alpha>5$ in $P(k)$, its fourth moment $\left\langle k^{4}\right\rangle $
is convergent and a mean-field critical behavior is obtained. When,
$3<\alpha\le5$, anomalous behavior of the thermodynamics quantities
is observed, due to divergence in $\left\langle k^{4}\right\rangle $,
but when $\alpha\le3$, the divergence is in the second moment, $\left\langle k^{2}\right\rangle $,
and the criticality vary with the size of the system, being an infinity
order phase transition in the thermodynamic limit. In the network
proposed in the Barabási-Albert (BA) model \citep{3}, the exponent
is limited by $\alpha=3$, and in previews of approximate and numerical
results, the infinity order phase transition is verified \citep{14,15}.
In addition to $\alpha=3$, the cases where $\left\langle k^{4}\right\rangle $
and $\left\langle k^{2}\right\rangle $ are convergent and divergent,
by Monte Carlo simulations \citep{16} and replica method \citep{17},
have been proven the non-trivial critical exponents, beyond the and
size-dependent critical temperature.

In these Monte Carlo simulations, the SFN, constructed with a selected
value of $\alpha$, is absent from the two growing mechanisms in the
BA model and created distributing degrees for the vertices based on
the exact value of $P(k)=Ak^{-\alpha}$ \citep{16}. For this exact
value is predefined the minimum $k_{0}$ and maximum $k_{m}$ degrees
of the network, and calculated the normalization constant of the distribution,
$A=\sum_{k=k_{0}}^{k_{m}}k^{\alpha}$. In this sense, supposing that
the number of sites with the degree $k_{m}$ is $N_{k_{m}}=1$, the
network size with these characteristics is given by $N=k_{m}^{\alpha}/A$.
From this way, fixing $k_{0}$ and varying the network sizes $N$,
Herrero \citep{16} could reproduce the main critical phenomena seen
analytically \citep{12,13}, by the FSS analysis of networks with
also non-fixed $k_{m}$.

In this work, we investigated the Ising model on a \emph{restricted}
SFN, where each site of the network is occupied by a spin variable
$\sigma=1/2$ that can assume values $\pm1$. Our network was built
for various integer values of the exponent $\alpha$, and divided
into two sublattices, each connection distributed by $P(k)$, a power-law
degree distribution, which should connect these sublattices. Besides
that, in a similar way to what was proposed by Herrero \citep{16}
to construct its random uncorrelated network, we also preview define
who will be $k_{0}$ and $k_{m}$ in our system. These values are
kept fixed while we vary the size $N$ of the network. Thereunto,
the minimal network size that we can use is defined by $N_{0}=k_{m}^{\alpha}/A$
and is not always obtained $N_{k_{m}}=1$. However, as we have the
same degrees on all network sizes, we always have a convergent $\left\langle k^{2}\right\rangle $
and $\left\langle k^{4}\right\rangle $, and consequently finite transitions
point, for ferromagnetic to paramagnetic phases, in the whole values
of $\alpha$. Thus here, through Monte Carlo simulations, we have
built phases diagrams of critical temperature $T_{c}$ as a function
of $k_{0}$ and $k_{m}$ for the studied values of $\alpha$, and
using the FSS theory, we have obtained the critical exponents for
the system.

This article is organized as follows: In Section \ref{sec:Model},
we describe the network used and the Hamiltonian model of the system.
In Section \ref{sec:Monte-Carlo-simulations}, we present the Monte
Carlo simulation method, some details concerning the simulation procedures
and the thermodynamic quantities of the system also necessary for
the application of FSS analysis. The behavior of thermodynamic quantities,
phases diagrams, and critical exponents are described in Section \ref{sec:Results}.
Finally, in Section \ref{sec:Conclusions}, we present our conclusions.

\section{Model \label{sec:Model}}

\begin{figure}
\begin{centering}
\includegraphics[bb=120bp 50bp 750bp 600bp,clip,scale=0.35]{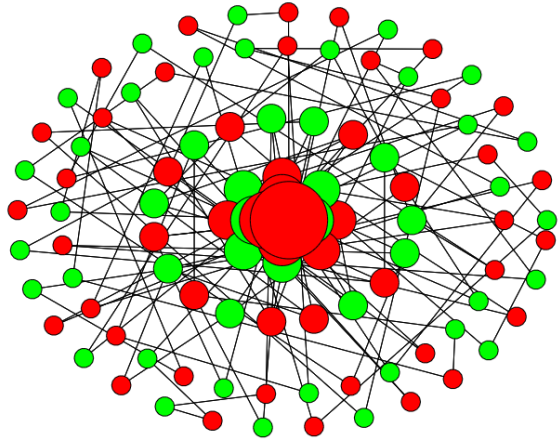}
\par\end{centering}
\caption{{\footnotesize{}Schematic representation of the }\emph{\footnotesize{}restricted}{\footnotesize{}
SFN. The red circles indicate the sites on one of the sublattices,
the green circles are the sites on the other sublattice, and the black
solid lines are the connections between the two sublattices. The size
of the circles is proportional to the sites degree, varying from $k_{0}=2$
to $k_{m}=8$, with $\alpha=3$, and $N=10^{2}$. \label{fig:1}}}
\end{figure}

The Ising model studied in this work have $N=L^{2}$ spins $\sigma_{i}=\pm1$,
on a \emph{restricted} SFN and a ferromagnetic interaction of strength
$J_{ik}$. To distribute the connections on this network, we first
defined $k_{0}$, $k_{m}$ and $\alpha$, i.e., the minimum and maximum
degree that the network is required to have and its exponent on the
distribution, respectively. Next, we calculated the normalization
constant of the distribution, $A=\sum_{k=k_{0}}^{k_{m}}k^{\alpha}$,
and found the smallest network that we will use in the system, $N_{0}=k_{m}^{\alpha}/A$.
With these values, we create a set of site numbers, $\{N_{k}\}$,
that will have the respective degrees $k$, $N_{k}=AN/k^{\alpha}$,
and distribute them by the network. For this distribution of connections,
we have divided the network into two sublattices, where one sublattice
plays the role of central spins, while the other sublattice contains
the spins in which the central spins can connect. Thus, starting with
the lowest degree, we select one site $i$ on the network, and its
sublattice will be the sublattice of central spins, then, from the
other sublattice, we select one site $j$ at random, that has not
yet received their respective connections. With this, we couple $j$
to the neighbors of $i$, and for the site $j$ we couple the site
$i$ to their neighbors. This process is done until $i$ has their
$k$ connections and visited all the set $\{N_{k}\}$. It is valid
to say that, for this network, there is not need to create these sublattices
and the connections can be created completely randomly between free
sites, but here, this implementation was done as a way to prepare
the system for future works in non-equilibrium systems without loss
of generality.

\begin{figure}
\begin{centering}
\includegraphics[scale=0.5]{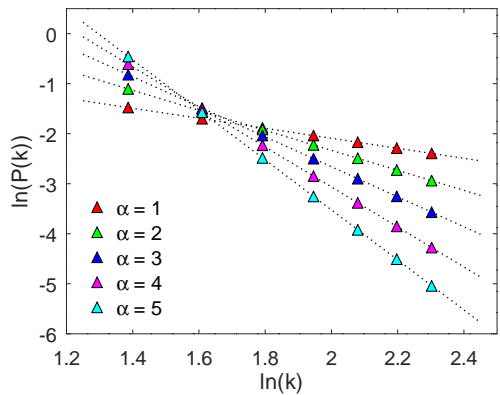}
\par\end{centering}
\caption{{\footnotesize{}Log-log plot of the degree distribution on }\emph{\footnotesize{}restricted}{\footnotesize{}
SFN for some selected values of $\alpha$, as shown in the figure.
All curves refer to the network size $N=256^{2}$, minimum degree
$k_{0}=4$ and maximum degree $k_{m}=10$. The dotted lines have the
exact expected slopes, $\alpha$, and the errors of the slopes are
in order of $10^{-4}$and the error bars are the same size or smaller
than the symbols. \label{fig:2}}}

\end{figure}

In Fig. \ref{fig:1}, we displayed an example of the network, with
$\alpha=3$, $k_{0}=2$, $k_{m}=8$ and $N=10^{2}$. The sites in
the middle of the figure are the more connected, while the peripheral
sites are the less connected, and sites from the sublattice green
are only connected with sites from the sublattice red.

Based on this construction, throughout work, we have selected the
integer values of $1\le\alpha\le5$, and network sizes $(32)^{2}\le N\le(256)^{2}$
to study the Ising model. In Fig. \ref{fig:2}, we shown the degree
distribution in the largest network size of the system for these selected
values of $\alpha$. With these distributions in the log-log plot,
we can see that the construction method used here, guarantees the
power-law form, as predicted in SFN.

The ferromagnetic Ising spin energy is described by the Hamiltonian
on the form

\begin{equation}
\mathcal{H}=-\sum_{\left\langle i,j\right\rangle }J_{ij}\sigma_{i}\sigma_{j}\label{eq:1}
\end{equation}
where the sum is over all pair of spins, and $J_{ij}$ is the ferromagnetic
interaction, and assuming the value of unity if sites $i$ and $j$
are connected by a link.

\section{Monte Carlo simulations \label{sec:Monte-Carlo-simulations}}

In the simulation of the system specified by the Hamiltonian in Eq.
(\ref{eq:1}) and with a \emph{restricted} SFN, we have chosen the
initial state of the system with all spins in the random states, and
a new configuration is generated by a Markov process. In this process,
for a given temperature $T$, exponents $\alpha$ of the degree distribution,
network size $N$, and minimum $k_{0}$ and maximum $k_{m}$ degree,
we choose at random a spin $\sigma_{i}$ on the network and change
its state by the one-spin flip mechanism with a transition rate given
by the following Metropolis prescription

\begin{equation}
W(\sigma_{i}\to\sigma_{i}^{\prime})=\left\{ \begin{array}{cccc}
e^{\left(-\Delta E/k_{B}T\right)} & \textrm{if} & \Delta E>0\\
1 & \textrm{if} & \Delta E\le0 & ,
\end{array}\right.\label{eq:2}
\end{equation}
where $\Delta E$ is the change in energy after flipping the spin,
$\sigma_{i}\to\sigma_{i}^{\prime}$, $k_{B}$ is the Boltzmann constant,
and $T$ the temperature of the system. Therefore, in this scenario,
the acceptance of a new state is made if $\Delta E\le0$, but, in
the case where $\Delta E>0$, the acceptance is pondered by the probability
$\exp\left(-\Delta E/k_{B}T\right)$ and just it is accepted by choosing
a random number $0<\xi<1$, where $\xi\le\exp\left(-\Delta E/k_{B}T\right)$.
On the other hand, if no one of these conditions was satisfied, we
do not change the state of the spin. 

Repeating the Markov process $N$ times, we have a Monte Carlo Step
(MCS). In our simulations, we have waited $10^{4}$ MCS for the system
reach the equilibrium state, in all lattice sizes and value of the
parameters. To calculate the thermal averages of the interest quantities,
we used more $4\times10^{4}$ MCS. The average over samples were done
using $10$ independent samples for any configuration. 

After reaching the equilibrium state, we have measured the following
thermodynamic quantities: magnetization per spin $\textrm{m}_{\textrm{L}}$,
magnetic susceptibility $\textrm{\ensuremath{\chi}}_{\textrm{L}}$
and reduced fourth-order Binder cumulant $\textrm{U}_{\textrm{L}}$:

\begin{equation}
\textrm{m}_{\textrm{L}}=\frac{1}{N}\left[\left\langle \sum_{i=1}^{N}\sigma_{i}\right\rangle \right],\label{eq:3}
\end{equation}

\begin{equation}
\textrm{\ensuremath{\chi}}_{\textrm{L}}=\frac{N}{k_{B}T}\left[\left\langle \textrm{m}_{\textrm{L}}^{2}\right\rangle -\left\langle \textrm{m}_{\textrm{L}}\right\rangle ^{2}\right],\label{eq:4}
\end{equation}

\begin{equation}
\textrm{U}_{\textrm{L}}=1-\frac{\left[\left\langle \textrm{m}_{\textrm{L}}^{4}\right\rangle \right]}{3\left[\left\langle \textrm{m}_{\textrm{L}}^{2}\right\rangle ^{2}\right]},\label{eq:5}
\end{equation}
where $\left[\ldots\right]$ represents the average over the samples,
and $\left\langle \ldots\right\rangle $ the thermal average over
the MCS in the equilibrium state. In the vicinity of the critical
temperature $T_{c}$, the above-defined quantities obey the following
finite-size scaling relations:

\begin{equation}
\textrm{m}_{\textrm{L}}=L^{-\beta/\nu}m_{0}(L^{1/\nu}\epsilon),\label{eq:6}
\end{equation}

\begin{equation}
\textrm{\ensuremath{\chi}}_{\textrm{L}}=L^{\gamma/\nu}\chi_{0}(L^{1/\nu}\epsilon),\label{eq:7}
\end{equation}

\begin{equation}
\textrm{U}_{\textrm{L}}^{\prime}=L^{1/\nu}\frac{U_{0}^{\prime}(L^{1/\nu}\epsilon)}{T_{c}},\label{eq:8}
\end{equation}
where $\epsilon=(T-T_{c})/T_{c}$, $m_{0}(L^{1/\nu}\epsilon)$, $\chi_{0}(L^{1/\nu}\epsilon)$
and $U_{0}(L^{1/\nu}\epsilon)$ are the scaling functions, and $\beta$,
$\gamma$ and $\nu$ are the magnetization, magnetic susceptibility
and length correlation critical exponents, respectively.

Using the Eqs. (\ref{eq:6}), (\ref{eq:7}), (\ref{eq:8}) and the
data from simulations for the network sizes $32\le L\le256$, we have
obtained the critical exponents ratios, $\beta/\nu$, $\gamma/\nu$
and $1/\nu$ from the slope of $\textrm{m}_{\textrm{L}}(T_{c})$,
$\textrm{\ensuremath{\chi}}_{\textrm{L}}(T_{c})$ and $\textrm{U}_{\textrm{L}}^{\prime}(T_{c})$
as a function of $L$ in a log-log plot. Aside from that, we also
used data collapse from scaling functions to estimate the critical
exponent values.

\section{Results \label{sec:Results}}

\begin{figure}
\begin{centering}
\includegraphics[scale=0.45]{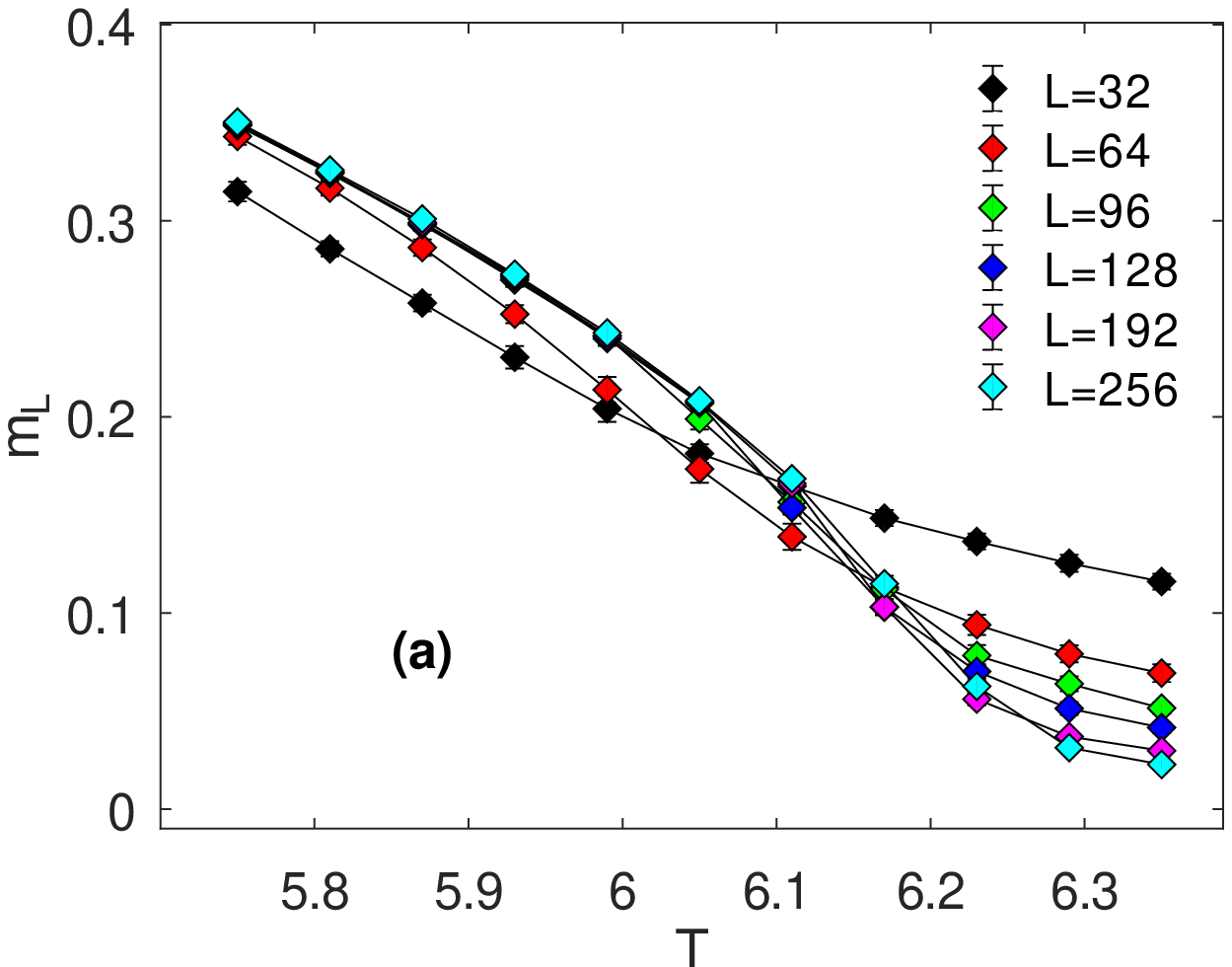}\hspace{0.1cm}\includegraphics[scale=0.45]{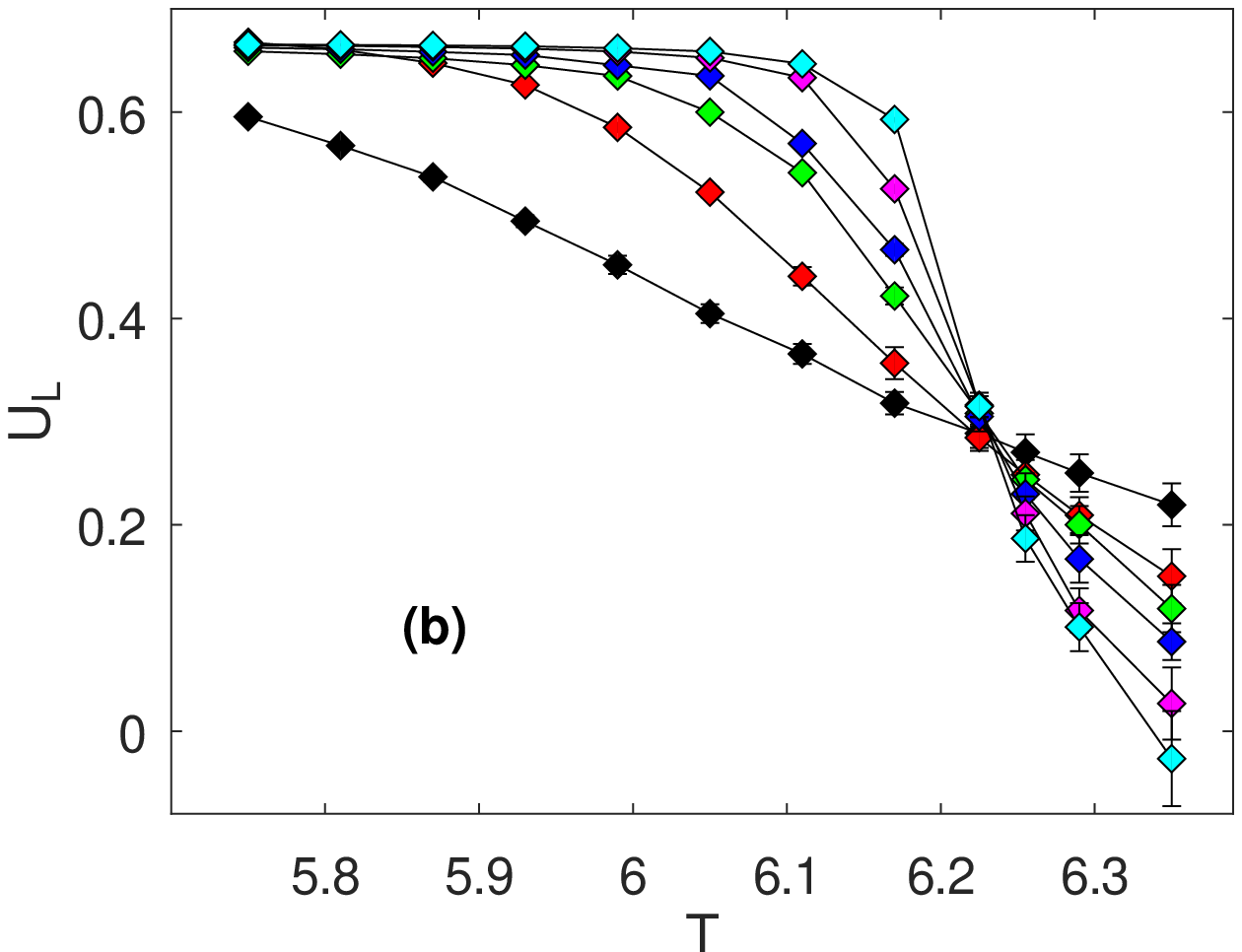}\hspace{0.1cm}\includegraphics[scale=0.45]{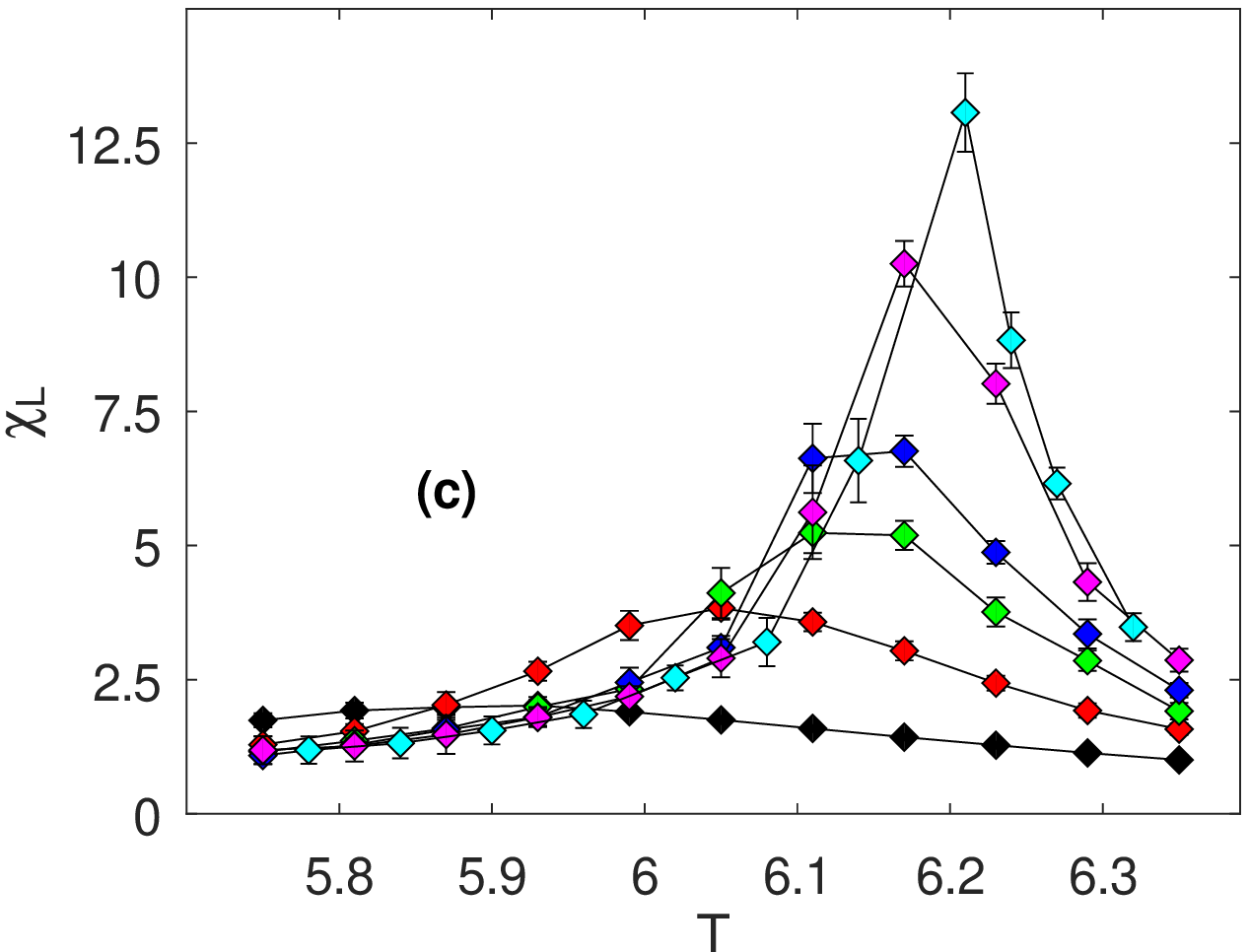}
\par\end{centering}
\caption{{\footnotesize{}Thermodynamic quantities as a function of temperature
$T$ for a fixed value of $\alpha=1$, $k_{0}=4$ and $k_{m}=10$,
and for different network sizes, as present in the figure. (a) Magnetization
$\textrm{m}_{\textrm{L}}$, (b) reduced fourth-order Binder cumulant
$\textrm{U}_{\textrm{L}}$, and (c) susceptibility $\textrm{\ensuremath{\chi}}_{\textrm{L}}$.
\label{fig:3}}}
\end{figure}

The interesting results about the critical phenomena in complex networks,
more specifically on random uncorrelated networks, led us to understand
the importance of its degree distribution and respective moments.
The Ising model on the uncorrelated SFN, in the limit where $N\to\infty$,
$k_{m}\to\infty$ and the changing in its structure for $\alpha>3$,
decreases the number of more connected spins, admitting a finite-order
phase transition until reaches the standard mean-field critical behavior
due to strong correlations of most connected vertices in their neighborhood
\citep{12,13}. However, here, by the restriction of maximum $k_{m}$
and minimum $k_{0}$ degree, and minimum network size, when $N\to\infty$,
$k_{m}$ keeps finite, consequently changing the number of more and
less connected sites, their correlations, and the critical phenomena
as we can see in our results.

\begin{figure}
\begin{centering}
\includegraphics[scale=0.5]{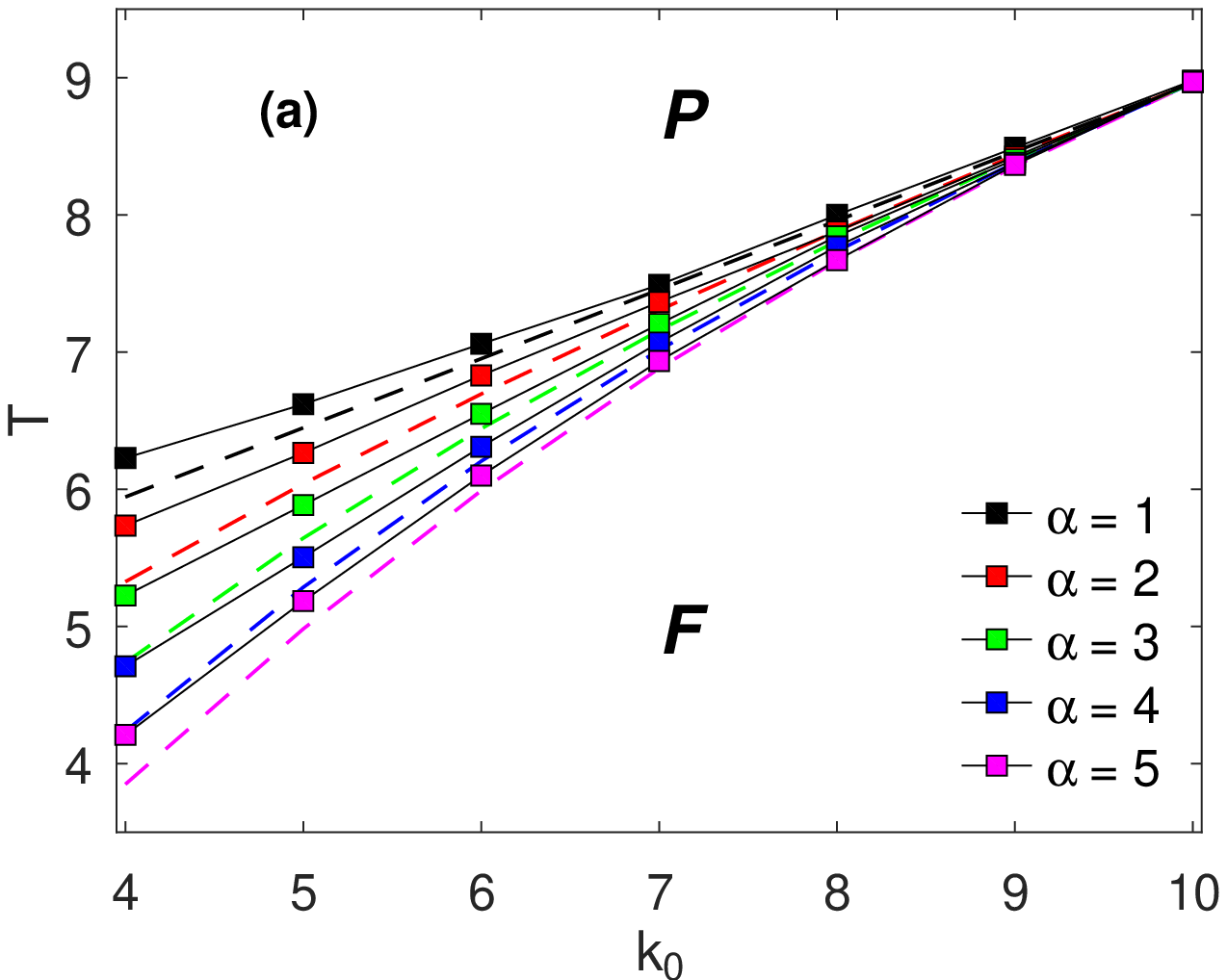}\hspace{0.25cm}\includegraphics[scale=0.5]{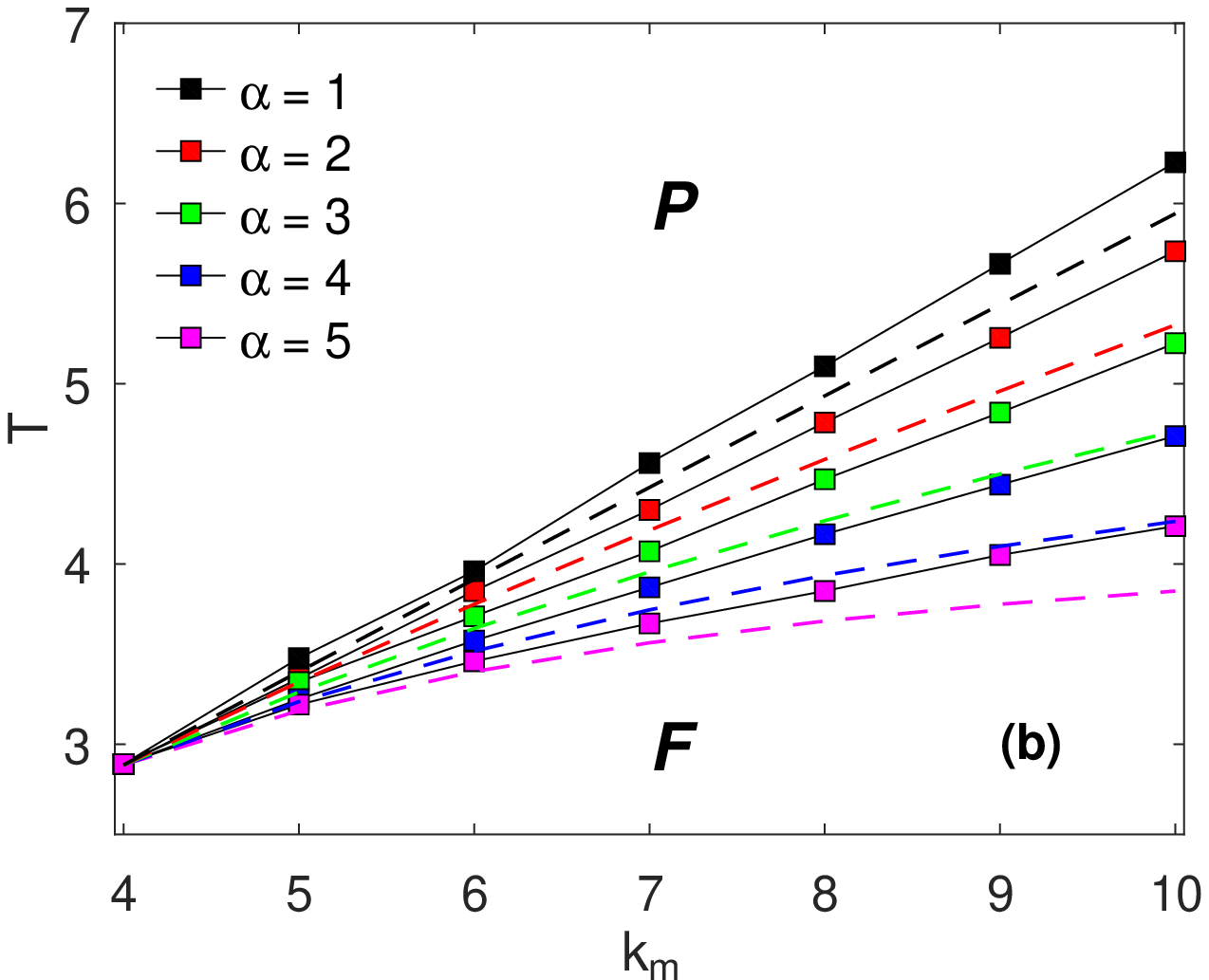}
\par\end{centering}
\caption{{\footnotesize{}(a) Phase diagram of temperature $T$ as a function
of $k_{0}$ for a fixed value of $k_{m}=10$ and (b) as a function
of $k_{m}$ for a fixed value of $k_{0}=4$. Both figures representing
transitions between $F$ to $P$ phases. The solid lines are just
guide for the eyes and the dashed lines represent the analytical result
obtained by Eq. (\ref{eq:9}). \label{fig:4}}}
\end{figure}

To begin with, we have identified the point of transition between
the ferromagnetic $F$ to the paramagnetic $P$ phases through the
curves of the fourth-order Binder cumulant $\textrm{U}_{\textrm{L}}$
with different network sizes \citep{18,19,20,21}. This critical point
and second-order phase transition can be identified by the crossing
point of $\textrm{U}_{\textrm{L}}$ curves, and an example is shown
in Fig. \ref{fig:3}. In this figure, we have presented as one of
the best results for thermodynamic quantities obtained by Eqs. (\ref{eq:3}),
(\ref{eq:4}) and (\ref{eq:5}), being that in Fig. \ref{fig:3}(a),
we can see the behavior of the magnetization $\textrm{m}_{\textrm{L}}$
as a function of $T$, the reduced fourth-order Binder cumulant $\textrm{U}_{\textrm{L}}$
(see Fig. \ref{fig:3}(b)), and the magnetic susceptibility $\textrm{\ensuremath{\chi}}_{\textrm{L}}$
(see Fig. \ref{fig:3}(c)). In this case, we have used a fixed value
of $\alpha=1$, $k_{0}=4$ and $k_{m}=10$, and different network
sizes, as present in Fig. \ref{fig:3}.

\begin{figure}
\begin{centering}
\includegraphics[scale=0.5]{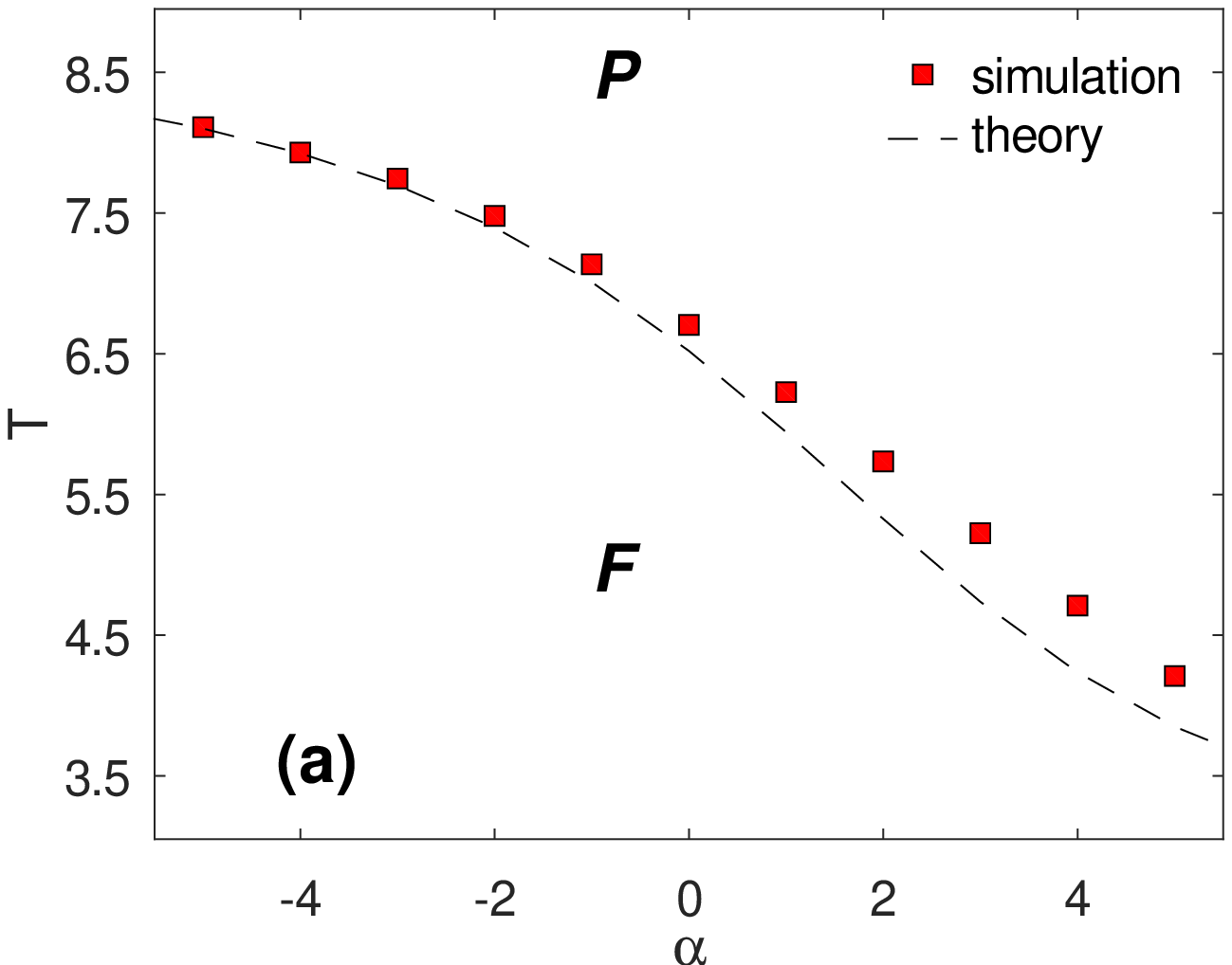}\hspace{0.25cm}\includegraphics[scale=0.5]{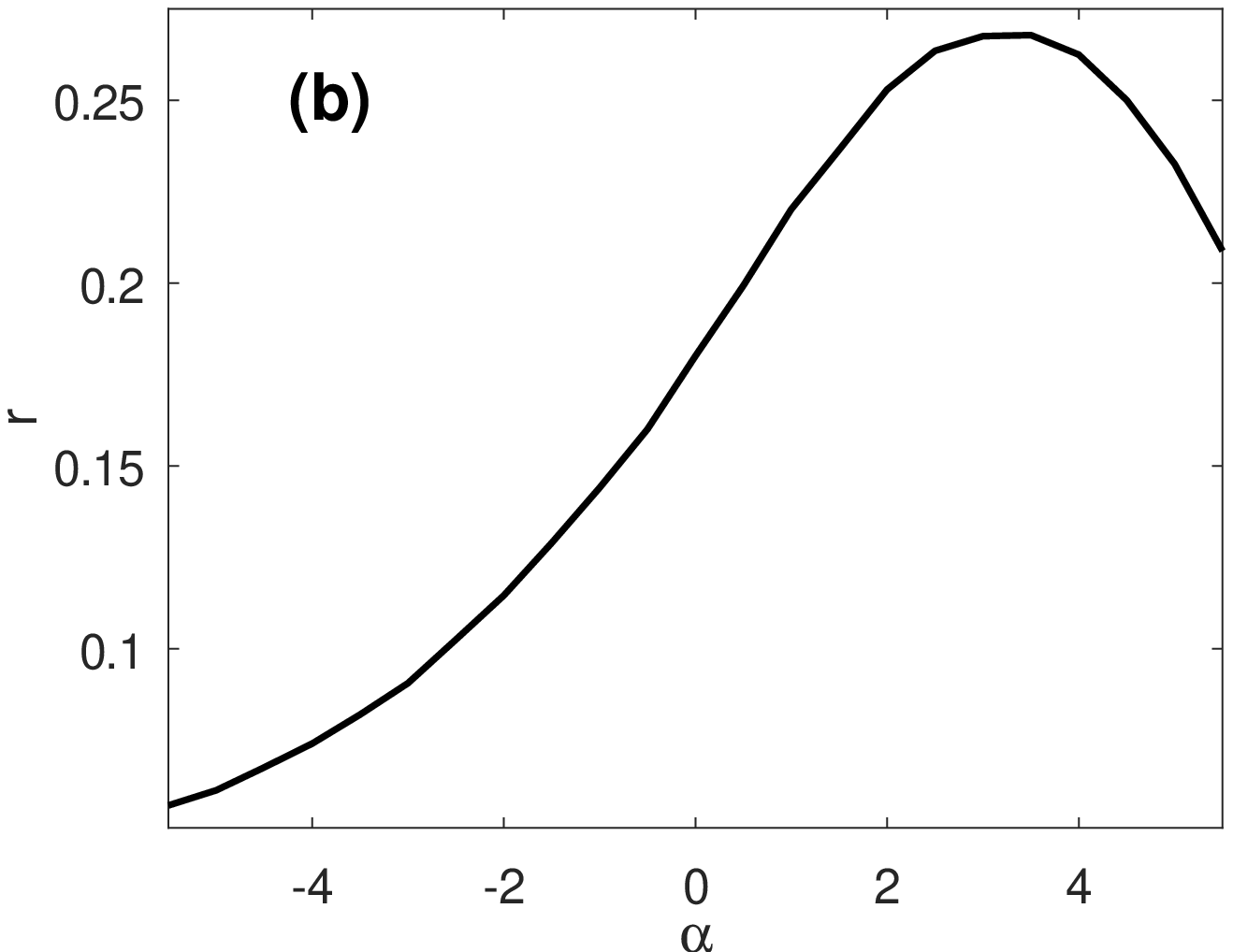}
\par\end{centering}
\caption{{\footnotesize{}(a) Phase diagram of temperature $T$ as a function
of $\alpha$, for the transition between $F$ to $P$ phases, with
fixed values of $k_{0}=4$ and $k_{m}=10$. The dashed line is the
theoretical result expected for random uncorrelated networks and obtained
using the Eq. (\ref{eq:9}). The red squares are the simulation data
points on the }\emph{\footnotesize{}restricted}{\footnotesize{} SFN,
and calculated by the crossing of} $\textrm{U}_{\textrm{L}}${\footnotesize{}
curves. (b) Degree-degree correlations $r$ as a function of $\alpha$
and obtined by the Eq. \ref{eq:10}. \label{fig:5}}}
\end{figure}

With the critical points obtained, the phase diagrams were built,
see Fig. \ref{fig:4}. For these diagrams, the temperature as a function
of $k_{0}$, $k_{m}$ and $\alpha$ is studied, and for that, we have
used $4\le k_{0},k_{m}\le10$ and integer values of $\alpha$ . The
continuous transition lines can be seen in Fig. \ref{fig:4}(a) for
temperature $T$ versus $k_{0}$, and for a fixed value of $k_{m}=10$,
and some selected values of $\alpha$, where we can verify that have
a finite critical point for all the values of $\alpha$. Because we
have a fixed value of $k_{m}$, as we increase $k_{0}$ the number
of different degrees on the network decrease until the whole system
has the coordination number $k_{0}=k_{m}=10$, and when that happens,
the exponent of the degree distribution is no longer important, returning
us to a unique critical point for all the exponents $\alpha$. On
the other hand, in the case where $k_{0}\ne k_{m}$, the degrees can
be distributed on the network, and for different values of $\alpha$
is showed the presence of different critical points in the system,
being that each value of $\alpha$ is responsible for a distinct curve
in Fig. \ref{fig:4}(a). Instead, we can fix $k_{0}=4$ and varying
$k_{m}$, and as can seen in Fig. \ref{fig:4}(b) for $T$ versus
$k_{m}$. Thus, we can observe the presence of the same qualitative
behavior, like, continuous transitions lines for the whole parameter
values, distinct curves for distinct exponents $\alpha$, and a unique
critical point when $k_{m}=k_{0}$. Despite these similarities, if
we approximate these curves by a linear fit, the slopes of curves
in Fig. \ref{fig:4}(a) have an increasing behavior according we increase
$\alpha$, while the slopes of curves in Fig. \ref{fig:4}(b) decrease
according to increase $\alpha$. This is visually perceptible, but
also is explicit in the variables $\Theta_{k_{0}}$ and $\Theta_{k_{m}}$
presented in Tab. \ref{tab:1}.

\begin{table}
\begin{centering}
\begin{tabular}{|c|c|c|c|c|c|}
\hline 
{\footnotesize{}$\alpha$} & {\footnotesize{}$\Theta_{k_{0}}$} & {\footnotesize{}$\Theta_{k_{m}}$} & {\footnotesize{}$-\beta/\nu$} & {\footnotesize{}$\gamma/\nu$} & {\footnotesize{}$1/\nu$}\tabularnewline
\hline 
\hline 
{\footnotesize{}$1$} & {\footnotesize{}$0.46\pm0.04$} & {\footnotesize{}$0.55\pm0.04$} & {\footnotesize{}$0.45\pm0.04$} & {\footnotesize{}$1.04\pm0.05$} & {\footnotesize{}$0.99\pm0.05$}\tabularnewline
\hline 
{\footnotesize{}$2$} & {\footnotesize{}$0.54\pm0.02$} & {\footnotesize{}$0.47\pm0.01$} & {\footnotesize{}$0.53\pm0.04$} & {\footnotesize{}$0.92\pm0.06$} & {\footnotesize{}$0.95\pm0.04$}\tabularnewline
\hline 
{\footnotesize{}$3$} & {\footnotesize{}$0.62\pm0.04$} & {\footnotesize{}$0.38\pm0.03$} & {\footnotesize{}$0.52\pm0.04$} & {\footnotesize{}$0.91\pm0.05$} & {\footnotesize{}$0.97\pm0.06$}\tabularnewline
\hline 
{\footnotesize{}$4$} & {\footnotesize{}$0.71\pm0.05$} & {\footnotesize{}$0.30\pm0.03$} & {\footnotesize{}$0.51\pm0.06$} & {\footnotesize{}$0.90\pm0.07$} & {\footnotesize{}$0.96\pm0.05$}\tabularnewline
\hline 
{\footnotesize{}$5$} & {\footnotesize{}$0.79\pm0.08$} & {\footnotesize{}$0.21\pm0.06$} & {\footnotesize{}$0.52\pm0.07$} & {\footnotesize{}$0.90\pm0.08$} & {\footnotesize{}$0.95\pm0.04$}\tabularnewline
\hline 
\end{tabular}
\par\end{centering}
\caption{{\footnotesize{}Slopes of curves presented in Fig. \ref{fig:4}(a),
$\Theta_{k_{0}}$, and Fig. \ref{fig:4}(b), $\Theta_{k_{m}}$, and
the ratio between the critical exponents obtained by the method presented
in Fig. \ref{fig:6}. \label{tab:1}}}
\end{table}

When $\alpha>3$, the analytical calculations predict well-defined
finite critical points in the $F$ to $P$ phase transition \citep{12,13,17}.
As a matter of comparison, the analytical results for the random uncorrelated
networks in which is known its first and second moment of degree distribution,
and introduced by equation 
\begin{equation}
T_{c}=\frac{2}{\ln\left(\frac{\left\langle k^{2}\right\rangle }{\left\langle k^{2}\right\rangle -2\left\langle k\right\rangle }\right)},\label{eq:9}
\end{equation}
also were plotted in Fig. \ref{fig:4} (see dashed lines). As we can
see in this figure, we also have plotted values of $\alpha<3$, and
its possible because instead of the usual integral approximation of
the sum in $\left\langle k\right\rangle $ and $\left\langle k^{2}\right\rangle $,
where result in $T_{c}\to0$ and $T_{c}\to\infty$, with $\alpha=2$
and $\alpha=3$, respectively, we have kept the sum, once that $k_{0}$
and $k_{m}$ are restricted. The dashed lines on Figs. \ref{fig:4}(a)
and \ref{fig:4}(b) are described by Eq. (\ref{eq:9}), and apparently,
we can observe the same signature, both in the analytical result and
simulations. However, in the building process of our \emph{restricted}
SFN, more connected sites are the last to be chosen to add their connections,
and their missing connections only can be attached to sites that were
not chosen yet, i.e., we have implicitly an increase of the degree-degree
correlations \citep{22,23}, once in the last steps of building the
network, more connected sites only can connect with more connected
sites. These correlations can be identified in the higher values of
$T_{c}$ in the simulations if compared with analytical results, because
as we increase the difference between $k_{0}$ and $k_{m}$, we also
increase the number of different degrees on the network and consequently
increase the possibility of degree-degree correlations. 

The degree-degree correlations were calculated here using the equation
\begin{equation}
r=\frac{M^{-1}\sum_{i}u_{i}v_{i}-\left[M^{-1}\sum_{i}\frac{1}{2}\left(u_{i}+v_{i}\right)\right]^{2}}{M^{-1}\sum_{i}\frac{1}{2}\left(u_{i}^{2}+v_{i}^{2}\right)-\left[M^{-1}\sum_{i}\frac{1}{2}\left(u_{i}+v_{i}\right)\right]^{2}},\label{eq:10}
\end{equation}
for a network with $M$ edges connecting the pair of vertices, and
their respective $u_{i}$ and $v_{i}$ degrees, as defined in Ref.
\citep{22}. These correlations are shown in Fig. \ref{fig:5}(b)
as a function of $\alpha$, and besides being independent of network
size, have a descending behavior according the number of more connected
sites increase, and ascending when this number become small, but relevant
yet. The peak of the correlation $r$ was observed in $\alpha=3.5$
and has value $r=0.268$. Therefore, the network presents a associative
mixing, confirming that sites with high degree, on the network, prefers
to connect on more connected sites, which is the case for high values
of $\alpha$. It is interesting to mention that many social networks
also have significant associative mixing, and do not present some
in complex network models, like random graphs or growing network model
(BA model \citep{22}).

Now that we have established the critical behavior of the system as
a function of the degrees on the network, using the most distinct
and distributed values of degree, $k_{0}=4$ and $k_{m}=10$, we will
effectively verify the critical behavior of the \emph{restricted}
SFN as a function of $\alpha$. As presented in Fig. \ref{fig:5}(a),
this verification was first done by using some points in Fig. \ref{fig:4}
by simulation data, and Eq. (\ref{eq:9}). By Eq. (\ref{eq:9}) is
explicit the lower values of $T_{c}$ where we have a high degree-degree
correlation, but, for small $r$, i.e., in lower values of $\alpha$,
the simulation results approach to analytical result of a random uncorrelated
network. That approach to analytical results is also resultant of
small influence of more connected sites, and also observed for $\alpha>3.5$.
By the fixed values of $k_{m}$ and $k_{0}$, the minimum and maximum
critical temperature values is limited to the case where $k_{m}=k_{0}=4$
on Fig. \ref{fig:4}(b) as we increase $\alpha$, because we will
never reach a network with coordination number $4$, without losing
the structure of degrees. The same is observable decreasing $\alpha$,
because we have the limit of $T_{c}$, in the case $k_{m}=k_{0}=10$
present in Fig. \ref{fig:4} (a) and unreachable in our \emph{restricted}
SFN. For some critical points in Fig. \ref{fig:5}, $1\le\alpha\le5$,
we have its explicit estimation and respective errors presented in
Tab. \ref{tab:2}.

Analytical results for the random uncorrelated networks also extend
to critical exponents in the Ising model, being that is predicted
for magnetic critical exponent, a mean-field character, $\beta=1/2$
for $\alpha>5$, with logarithmic corrections in $\alpha=5$, and
$\beta=1/(\alpha-3)$, and for $3<\alpha<5$. Interestingly, magnetic
susceptibility critical exponent, $\gamma$, has a mean-field universal
character for $\alpha>3$, i.e., $\gamma=1$ for the whole values
of the exponents of the degree distribution in which a finite order
phase transition is predicted, $\left\langle k^{2}\right\rangle <\infty$.
And in accordance with the scaling law relation $\gamma/\nu=2-\eta$,
is also expected a universal mean-field character for correlation
length, $\nu=1/2$, and Fisher exponent, $\eta=0$ \citep{24}. With
this results, here, only for $1\le\alpha\le5$ by limitation of the
network size and computational time, we have made a direct comparative
with this analytical results, using our \emph{restricted} SFN. Therefore,
we have calculated the critical exponents for the system, and this
was made by two methods. One of these methods, using the FSS relations,
is based on the slope of its linear fit, with the thermodynamic quantities
near to critical point, and the second one, is based on the data collapse
of thermodynamic quantities in the form of scaling functions \citep{18,19,20}.

\begin{figure}
\begin{centering}
\includegraphics[scale=0.45]{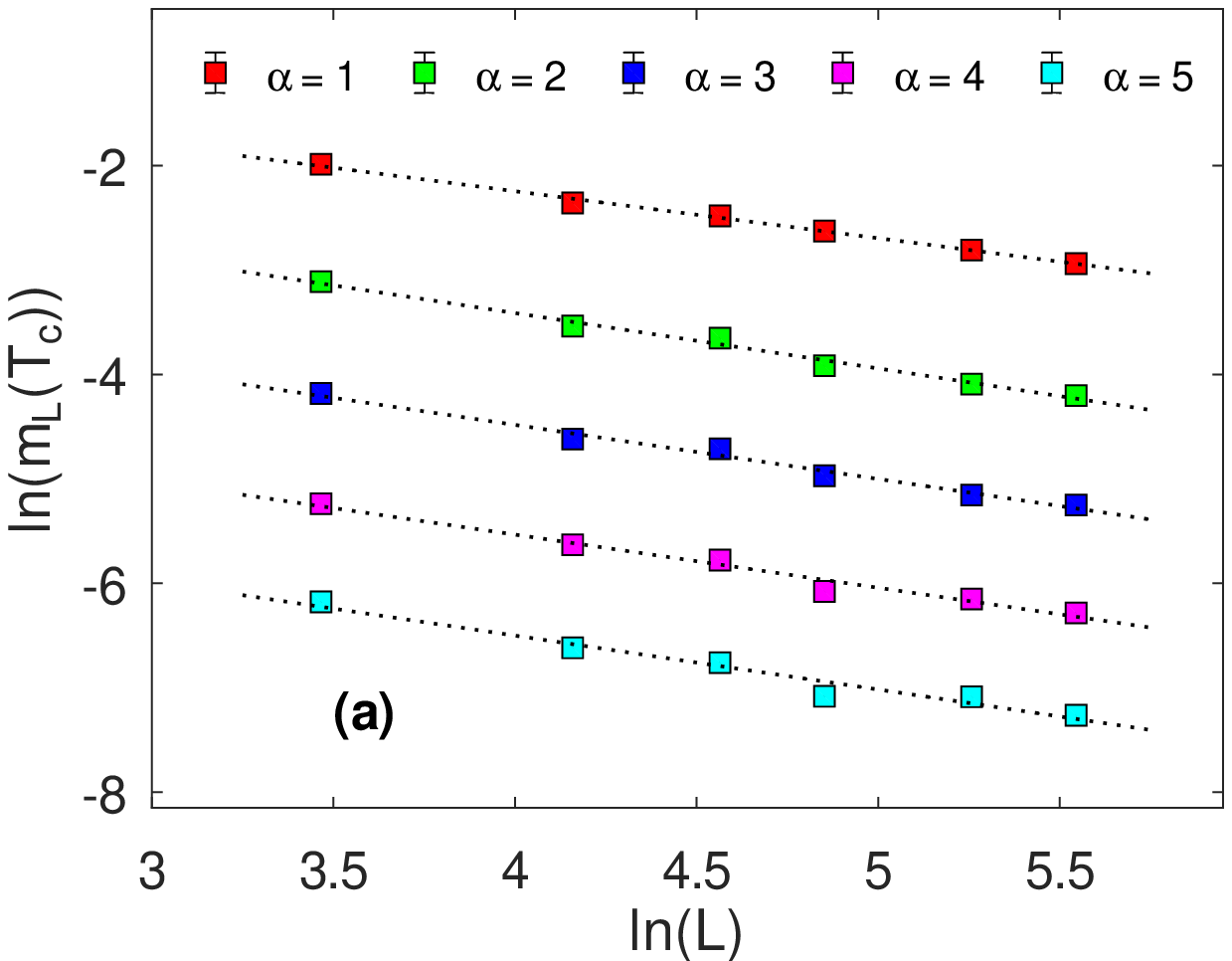}\hspace{0.1cm}\includegraphics[scale=0.45]{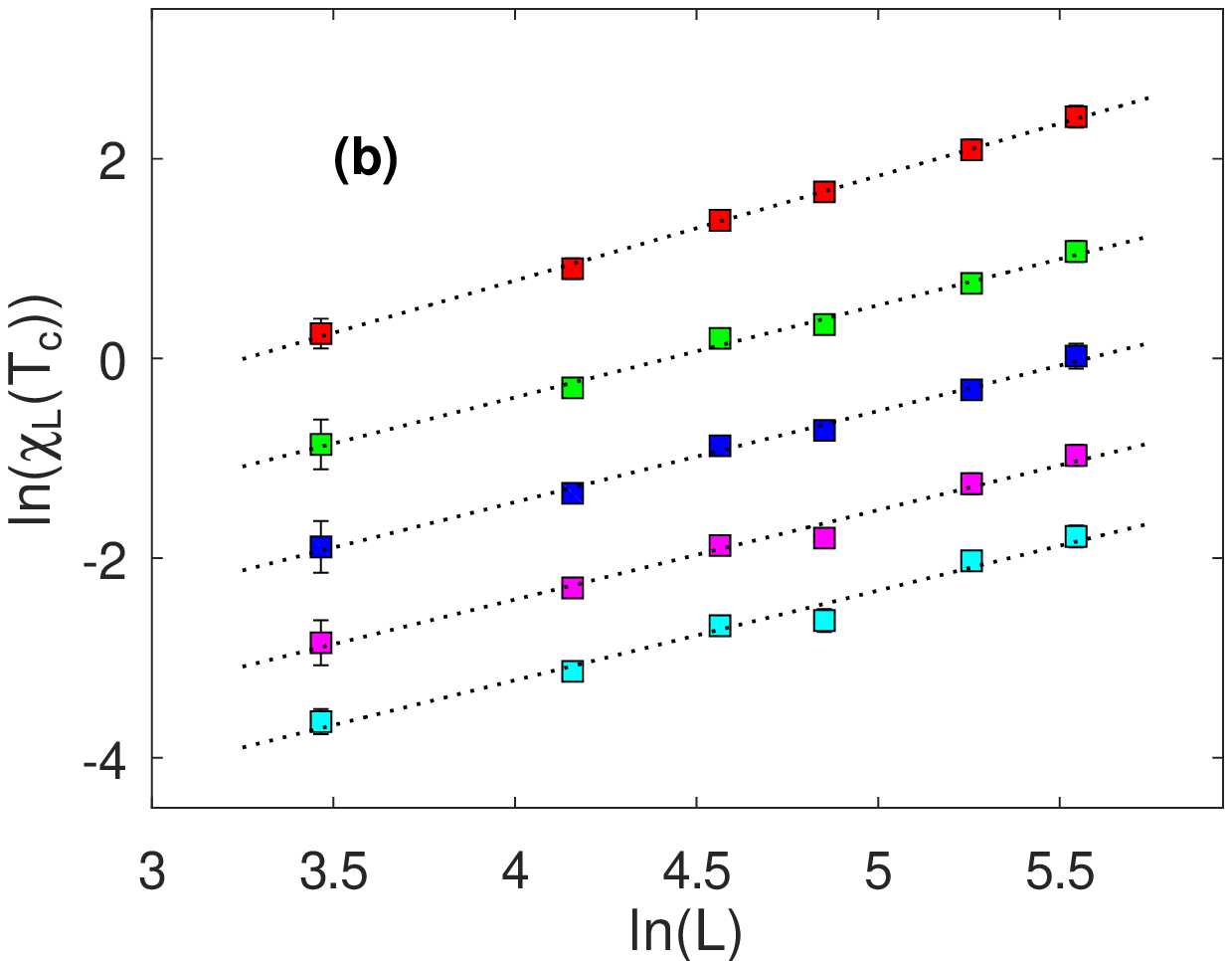}\hspace{0.1cm}\includegraphics[scale=0.45]{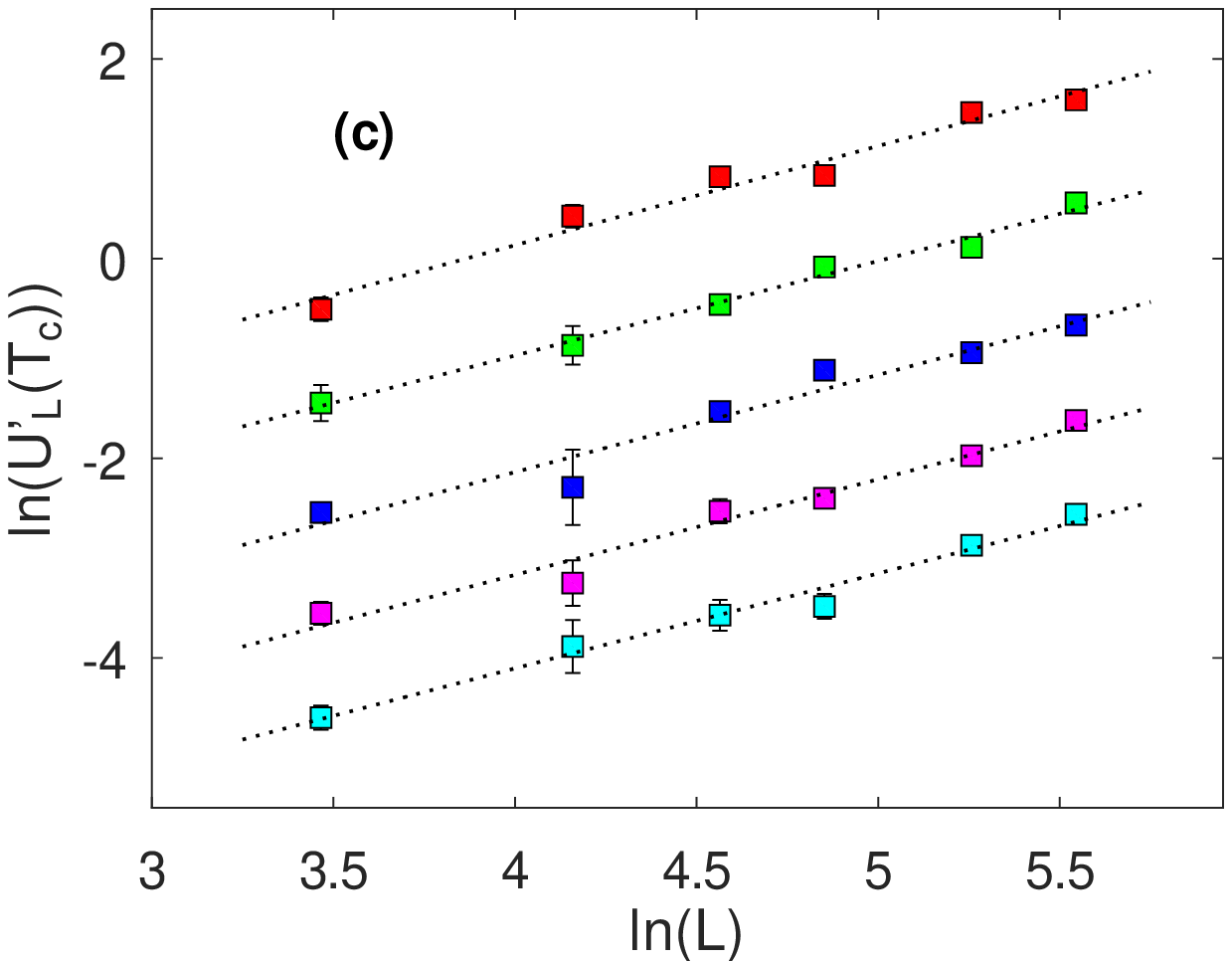}
\par\end{centering}
\caption{{\footnotesize{}Log-log plots of (a) the magnetization $\textrm{m}_{\textrm{L}}$,
(b) the susceptibility$\textrm{\ensuremath{\chi}}_{\textrm{L}}$,
and (c) the derivative of the cumulant$\textrm{U}_{\textrm{L}}^{\prime}$,
at the critical point, as a function of the effective length $L$.
We have fixed the values of $k_{0}=4$, $k_{m}=10$, and some selected
values $\alpha$, as indicated in the figure. As we are interest is
only on the slope, the linear coefficients are changed for better
visualization of each of the curves. The slopes obtained here can
be seen in Tab. \ref{tab:1}. \label{fig:6}}}
\end{figure}

We can find the critical exponents of the system by using two methods.
In the first method, we have used the data of thermodynamic quantities
near the critical point and for different network sizes. For instance,
in a log-log plot in Fig. \ref{fig:6}, we have fited the thermodynamic
quantities as a function of the effective length of the system $L$
and its slope returns us the ratios between the critical exponents
present in Eqs. (\ref{eq:6}), (\ref{eq:7}) and (\ref{eq:8}). For
the ratio $-\beta/\nu$ present in Eq. (\ref{eq:6}), the curves of
magnetization for different network sizes and values $\alpha$ is
presented in Fig. \ref{fig:6}(a), where we shown its linear fits.
In the same way, but now for the ratio $\gamma/\nu$, using the Eq.
(\ref{eq:7}), the linear fit of magnetic susceptibility curves can
be seen in Fig. \ref{fig:6}(b). Only these two ratios does not give
us all the critical exponents of the system. Therefore, we also have
used the forth-order Binder cumulant curves and its derivative, Eq.
(\ref{eq:8}), being that its linear fits is presented in Fig. \ref{fig:6}(c),
and give us the ratio $1/\nu$ that is the inverse of correlation
length exponent, and only then obtain the exponents $\beta$ and $\gamma$.
The set of exponents obtained by this method can be found in Tab.
\ref{tab:1}. However, it is necessary to emphasize that, because
we are dealing with a random network with modifiable degree distribution,
it is required to modify the Eqs. (\ref{eq:6}), (\ref{eq:7}) and
(\ref{eq:8}) to the mean-field scaling relations, in which we change
the effective length $L$ by the total number of spins on network,
$N$. This modification in the scaling relations can be easily implemented
in the obtained critical exponents, dividing $\nu$ by two, once that
$N=L^{2}$. 

\begin{figure}
\begin{centering}
\includegraphics[scale=0.5]{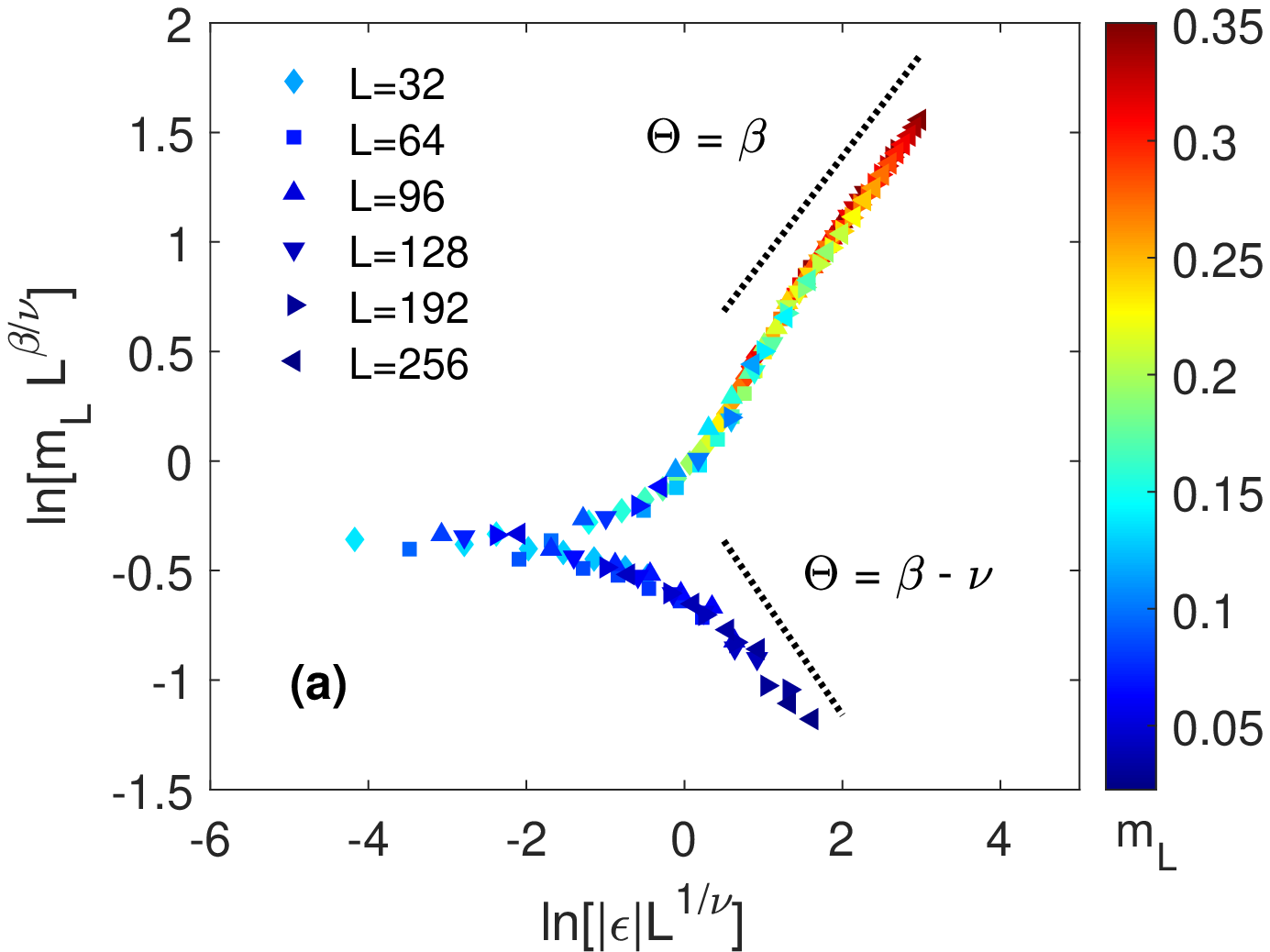}\hspace{0.25cm}\includegraphics[scale=0.5]{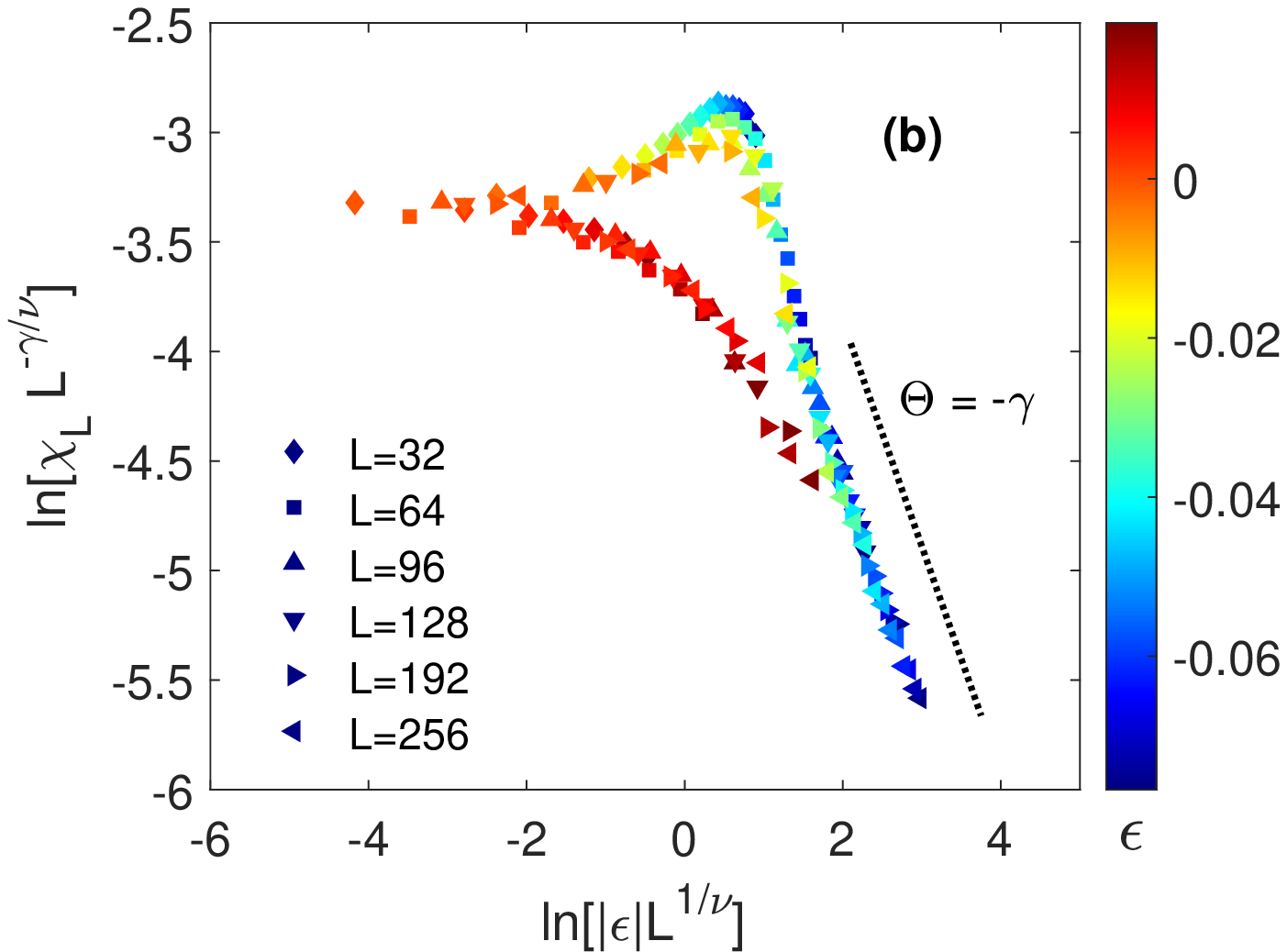}
\par\end{centering}
\caption{{\footnotesize{}Data collapse near the critical point for the magnetization
$\textrm{m}_{\textrm{L}}$ (a) and susceptibility $\textrm{\ensuremath{\chi}}_{\textrm{L}}$
(b) for different network sizes, as shown in the figure. Here, the
minimum and maximum degree are fixed $k_{0}=4$, $k_{m}=10$, and
$\alpha=1$. The log-log plots were used to obtain the slope $\Theta$
of scaling functions asymptotic behavior, i.e., distant of $\epsilon=0$.
The straight-dashed lines represent the asymptotic behavior of the
scaling functions, Eq. (\ref{eq:6}) and (\ref{eq:7}).\label{fig:7}}}
\end{figure}

\begin{table}
\begin{centering}
\begin{tabular}{|c|c|c|c|c|c|}
\hline 
{\footnotesize{}$\alpha$} & {\footnotesize{}$T_{c}$} & {\footnotesize{}$\beta$} & {\footnotesize{}$\gamma$} & {\footnotesize{}$\nu_{\textrm{m}_{\textrm{L}}}$} & {\footnotesize{}$\nu_{\textrm{\ensuremath{\chi}}_{\textrm{L}}}$}\tabularnewline
\hline 
\hline 
{\footnotesize{}$1$} & {\footnotesize{}$6.228\pm0.004$} & {\footnotesize{}$0.47\pm0.03$} & {\footnotesize{}$1.03\pm0.03$} & {\footnotesize{}$1.00\pm0.06$} & {\footnotesize{}$1.00\pm0.04$}\tabularnewline
\hline 
{\footnotesize{}$2$} & {\footnotesize{}$5.734\pm0.008$} & {\footnotesize{}$0.55\pm0.02$} & {\footnotesize{}$0.95\pm0.03$} & {\footnotesize{}$1.03\pm0.03$} & {\footnotesize{}$1.05\pm0.04$}\tabularnewline
\hline 
{\footnotesize{}$3$} & {\footnotesize{}$5.225\pm0.009$} & {\footnotesize{}$0.57\pm0.03$} & {\footnotesize{}$0.91\pm0.05$} & {\footnotesize{}$0.96\pm0.05$} & {\footnotesize{}$1.08\pm0.03$}\tabularnewline
\hline 
{\footnotesize{}$4$} & {\footnotesize{}$4.710\pm0.020$} & {\footnotesize{}$0.56\pm0.03$} & {\footnotesize{}$0.90\pm0.04$} & {\footnotesize{}$1.05\pm0.05$} & {\footnotesize{}$1.10\pm0.03$}\tabularnewline
\hline 
{\footnotesize{}$5$} & {\footnotesize{}$4.210\pm0.020$} & {\footnotesize{}$0.58\pm0.03$} & {\footnotesize{}$0.92\pm0.05$} & {\footnotesize{}$1.03\pm0.04$} & {\footnotesize{}$1.10\pm0.05$}\tabularnewline
\hline 
\end{tabular}
\par\end{centering}
\caption{{\footnotesize{}As a function of $\alpha$, here is showed critical
temperature $T_{c}$ present in the phases diagram of Fig. \ref{fig:5},
and critical exponents obtained by the data collapse method, being
that $\nu_{\textrm{m}_{\textrm{L}}}$ and $\nu_{\textrm{\ensuremath{\chi}}_{\textrm{L}}}$
are the correlation length exponent obtained by the magnetic and magnetic
susceptibility data collapsed curves, respectively. \label{tab:2}}}
\end{table}

In the second method, we can improve the values found for the critical
exponents, by collapsing the data points. Our main goal is to use
the thermodynamic quantities curves and different network sizes to
obtain the form of its scaling functions, as a function of $L^{1/\nu}\epsilon$.
It is possible because, in the proximity of the critical points, scaling
functions in Eqs. (\ref{eq:6}), (\ref{eq:7}) and (\ref{eq:8}) must
be independently of network sizes, if on its, is utilized the correct
critical exponents of the system, i.e., in the proximity of $T_{c}$
using the correct exponents on scaling relations we obtain a collapsed
curve in the form of scaling functions. In Fig. \ref{fig:7}, we show
some examples of the best collapsed curves obtained. In this figure,
what we have done is adjust the exponents on isolated scaling functions,
$m_{0}$ and $\chi_{0}$, as a function of $L^{1/\nu}\varepsilon$,
and when the thermodynamic quantity with different network size, best
collapses into a single curve, that exponents were considered the
critical exponents of the system. Fig. \ref{fig:7}(a) shown, for
$\alpha=1$, the collapsed curves of magnetization data, Eq. (\ref{eq:6}),
in which allowed us to adjust and obtain the exponents $\beta$ and
$\nu$, and besides that, in Fig. \ref{fig:7}(b) also for $\alpha=1$,
magnetic susceptibility data, Eq. (\ref{eq:7}), enable us to obtain
exponents $\gamma$ and $\nu$. That estimated critical exponents
for $1\le\alpha\le5$ is presented in Tab. \ref{tab:2}, but, we also
have to take into account the modification in scaling functions for
the mean-field character, and the real values of exponent $\nu$,
is only obtained dividing its results by two. 

\begin{figure}
\begin{centering}
\includegraphics[scale=0.5]{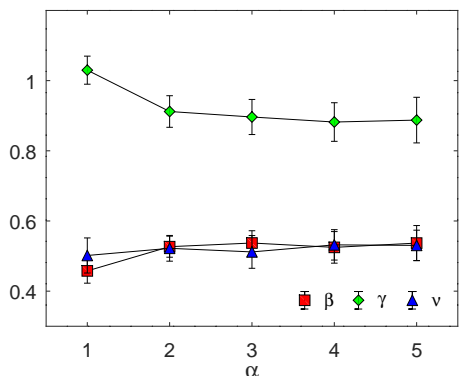}
\par\end{centering}
\caption{{\footnotesize{}Static critical exponents $\beta$, $\gamma$, and
$\nu$ as functions of the exponent $\alpha$. \label{fig:8}}}
\end{figure}

To the best comprehension of the critical exponents obtained for the
\emph{restricted} SFN, the average of its values, which is equivalent
in both methods utilized in this work, was plotted in Fig. \ref{fig:8}
as a function of $\alpha$. In lower values of $\alpha$, we tend
to a system with all the degrees having the same number of sites and
the mean-field critical behavior is present, but, when $\alpha$ increase,
the number of more connected sites decreases, and a slight deviation
from this behavior is observed. With that observations, we can say
that degree-degree correlations, also causes a deviation from the
expected mean-field behavior on random networks with degree distribution
convergent moments \citep{24,25,26}.

\section{Conclusions \label{sec:Conclusions}}

Here, we have employed Monte Carlo simulations to the study of thermodynamic
quantities and the critical behavior of the Ising model on a \emph{restricted}
SFN. When we fix the maximum $k_{m}$ and minimum $k_{0}$ number
of degrees for the whole network sizes, as we made on our \emph{restricted}
SFN, we always have convergent moments based in the degree distribution
$P(k)$. We have used a power-law degree distribution, and as the
analytical results predict, the convergent second and fourth moments
in the arbitrary degree distributions present a finite order phase
transition \citep{12,13,17,24}. We have obtained the critical points
of second-order phase transitions and built phase diagram of temperature
$T$ as a function of $k_{0}$ and $k_{m}$, and in which when we
have $k_{0}=k_{m}$ a random uncorrelated network is obtained. Therefore,
the critical points are in agreement with analytical calculations,
but, the increase on difference between $k_{0}$ and $k_{m}$, also
increase the degree-degree correlations and causes a deviation from
that calculations. The phase diagram of temperature $T$ as a function
of $\alpha$ also was built, where we always have a finite critical
temperature and a decrease in the critical point as we increase $\alpha$,
once that we also decrease the number of more connected sites on network.
With these critical points, we estimated the critical exponents for
the system as a function of $\alpha$, and different from what is
predicted by analytical results in a random uncorrelated SFN, here,
with lower values of $\alpha$, the increase in the number of more
connected sites reaches a mean-field critical behavior. Otherwise,
when $\alpha$ increase, a mean-field critical behavior is also observed,
but how are we dealing with correlated degrees on network, the decreasing
in the number of more connected sites causes a slight deviation on
that exponents. That mean-field behavior was predicted on networks
with convergent moments on its degree distribution and it was found
on a diversity of complex networks \citep{24,25,26}. In our work,
it was made on the Ising model with a power-law degree distributed
network and unexpected values of $\alpha$.

\end{document}